\documentclass[10pt,conference]{IEEEtran}
\IEEEoverridecommandlockouts

\usepackage{cite,url}
\usepackage{booktabs}
\usepackage{tikz}
\usepackage{pgfplots}
\usepackage{pgfplotstable}
\usepackage{amsmath,amssymb,amsfonts}
\usepackage{algorithmic}
\usepackage{graphicx}
\usepackage{caption}
\usepackage{subcaption}
\usepackage{textcomp}
\usepackage{amsmath,amsthm}
\usetikzlibrary{pgfplots.groupplots}
\usepackage{xcolor}
\usepackage[labelfont=bf]{caption}
\def\BibTeX{{\rm B\kern-.05em{\sc i\kern-.025em b}\kern-.08em
    T\kern-.1667em\lower.7ex\hbox{E}\kern-.125emX}}

\usepackage[linesnumbered,ruled]{algorithm2e}
\usepackage{amssymb}
\usepackage{colortbl}
\usepackage{tcolorbox}
\usepackage{listings}
\usepackage{makecell}
\usepackage{flushend}

\definecolor{dkgreen}{rgb}{0,0.6,0}
\definecolor{gray}{rgb}{0.5,0.5,0.5}
\definecolor{mauve}{rgb}{0.58,0,0.82}
\definecolor{orangep}{rgb}{0.71, 0.43, 0.89}
\definecolor{orp}{rgb}{1, 0.7, 0.278}

\definecolor{darkBlue}{rgb}{0.000000,0.000000,0.545098}
\definecolor{darkGreen}{rgb}{0.000000,0.392157,0.000000}
\definecolor{DarkGray}{gray}{0.4}
\definecolor{javared}{rgb}{0.6,0,0} 
\definecolor{javagreen}{rgb}{0.25,0.5,0.35} 
\definecolor{javapurple}{rgb}{0.5,0,0.35} 
\definecolor{javadocblue}{rgb}{0.25,0.35,0.75} 
\definecolor{lightgray}{gray}{0.95}
\definecolor{shadecolor}{RGB}{150,150,150}
\definecolor{blueA}{RGB}{204,229,255}
\definecolor{redA}{RGB}{112,0, 0}
\lstdefinestyle{MyCSmallStyle} {
  language=C++,
  frame=none,
  xleftmargin=15pt,
  stepnumber=1, 
  numbers=left, 
  numbersep=5pt,
  numberstyle=\tiny\color[gray]{0.177}, 
  belowcaptionskip=\bigskipamount,
  captionpos=b, 
  escapeinside={*'}{'*},
  tabsize=5,
  emphstyle={\bf},
  escapechar=!,
  basicstyle=\scriptsize\ttfamily,
  keywordstyle=\color{javapurple}\bfseries,
  stringstyle=\color{javared},
  commentstyle=\color{javagreen},
  morecomment=[s][\color{javadocblue}]{/**}{*/},
  showspaces=false,
  columns=flexible,
  showstringspaces=false,
  morecomment=[l]{//},
  tabsize=2,
  breaklines=true,
  moredelim=[is][\underbar]{^}{^}
}
\usepackage{graphicx}

\lstdefinelanguage{Scala}{
  keywords={typeof, new, true, false, catch,def,val, function, return, null, catch, switch, var, if, in, while, do, else, case, break, assert, static, void ,declare, const, for, define,fun, ite,class, not, check,sat,String, Int, ArrayList, foreach},
  keywordstyle=\color{javapurple}\bfseries,
  ndkeywords={ export,extends, boolean, throw, implements, import, this, abstract,reduceByKey, reduce, filter, map, reduceByKey, join, Join1, public, mapValues },
  ndkeywordstyle=\color{blue}\bfseries,
  otherkeywords={+, =>,<=, =, >,< , ||},
  identifierstyle=\color{black},
  sensitive=false,
  comment=[l]{//},
  morecomment=[s]{/*}{*/},
  commentstyle=\color{purple}\ttfamily,
  stringstyle=\color{red}\ttfamily,
  morestring=[b]',
  morestring=[b]"
}

\SetKwInput{KwInput}{Input}                
\SetKwInput{KwOutput}{Output}              

\newcommand{\randomfuzzS}{{\small{\sc{FuzzS}}}\xspace}
\newcommand{\randomfuzzVS}{{\small{\sc{FuzzVS}}}\xspace}

\newcommand{\tool}{{\small{\textsc{BigFuzz}}}\xspace}

\newcommand{\codefont}[1]{{\texttt{#1}}}
\newcommand{\eg}{\emph{e.g.,}\xspace}

\begin{document}



\title{\LARGE{Efficient Fuzz Testing for Apache Spark\\ Using Framework Abstraction}}
\makeatletter
\newcommand{\linebreakand}{%
  \end{@IEEEauthorhalign}
  \hfill\mbox{}\par
  \mbox{}\hfill\begin{@IEEEauthorhalign}
}
\makeatother

\author{
\IEEEauthorblockN{Qian Zhang}
\IEEEauthorblockA{University of California, Los Angeles}
\and
\IEEEauthorblockN{Jiyuan Wang}
\IEEEauthorblockA{University of California, Los Angeles}
\and
\IEEEauthorblockN{Muhammad Ali Gulzar}
\IEEEauthorblockA{Virginia Tech}
\linebreakand
\IEEEauthorblockN{Rohan Padhye}
\IEEEauthorblockA{Carnegie Mellon University}
\and
\IEEEauthorblockN{Miryung Kim}
\IEEEauthorblockA{University of California, Los Angeles}
}

\maketitle

\begin{abstract}
The emerging data-intensive applications are increasingly dependent on data-intensive scalable computing (DISC) systems, such as Apache Spark, to process large data. Despite their popularity, DISC applications are hard to test. In recent years, fuzz testing has been remarkably successful; however, it is nontrivial to apply such traditional fuzzing to big data analytics directly because: (1) the long latency of DISC systems prohibits the applicability of fuzzing, and (2) conventional branch coverage is unlikely to identify application logic from the DISC framework implementation. We devise a novel fuzz testing tool called {\tool} that automatically generates concrete data for an input Apache Spark program. The key essence of our approach is that we abstract the dataflow behavior of the DISC framework with executable specifications and we design schema-aware mutations based on common error types in DISC applications. Our experiments show that compared to random fuzzing, \tool is able to speed up the fuzzing time by 1477X, improves application code coverage by 271\%, and achieves 157\% improvement in detecting application errors. The demonstration video of \tool is available at \url{\footnotesize https://www.youtube.com/watch?v=YvYQISILQHs&feature=youtu.be}.
\end{abstract}


\section{Introduction}\label{sec:introduction}
The importance of emerging data-intensive applications continues to grow at an increasing rate. Data-intensive scalable computing (DISC) systems, such as Google's MapReduce~\cite{dean2004mapreduce}, Apache Hadoop~\cite{hadoop}, and Apache Spark~\cite{spark}, enable processing massive data sets by providing distributed, parallel versions of dataflow operator (e.g., \codefont{map}, \codefont{reduce}, \codefont{join}, etc.) implementations with application logic expressed in terms of user-defined functions (UDFs). Although DISC systems are becoming widely available to the industry, DISC applications are hard to test. The standard practice for testing such DISC applications today is to select a subset of inputs based on the developers' hunch with the hope that it will reveal possible defects. Not surprisingly, these applications are thus not tested thoroughly and may not be robust to bugs and failures in the production setting. 

In recent years, fuzz testing has emerged as an effective technique for testing software systems~\cite{afl,FuzzingSurvey}. 
The effectiveness of such fuzzing techniques is based on two inherent yet oversight assumptions: (1) it takes a minuscule amount of time in the order of milliseconds to execute the target application, and (2) a set of arbitrary input mutations is likely to yield meaningful inputs. However, our extensive experience with DISC applications suggests that neither of the two assumptions holds for big data analytics. 

\begin{figure*}
\centering
\hfill\begin{minipage}[c]{.30\textwidth}
\includegraphics[width=\textwidth]{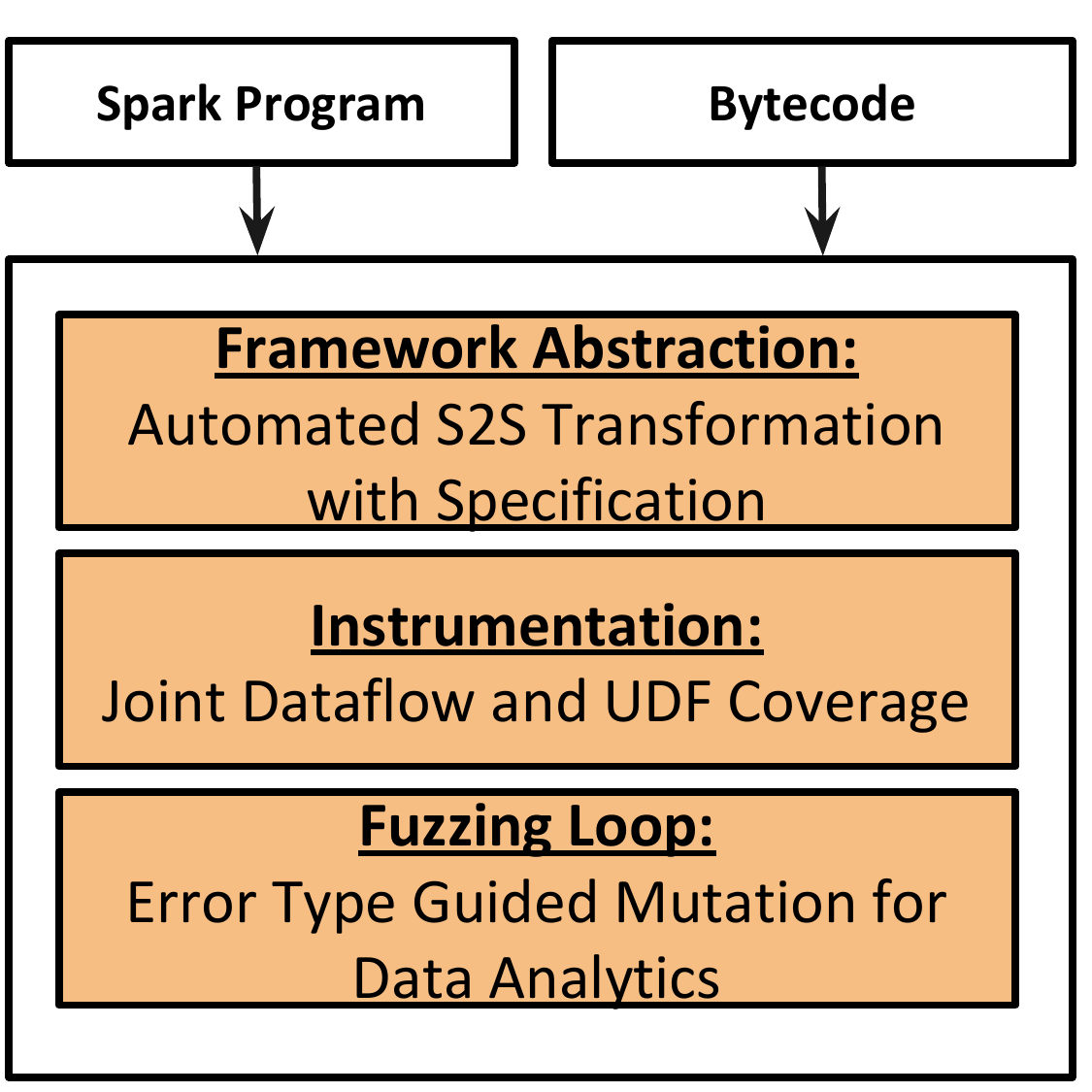}
\end{minipage} 
\hfill
\begin{minipage}[c]{.65\textwidth}
 \includegraphics[width=\textwidth]{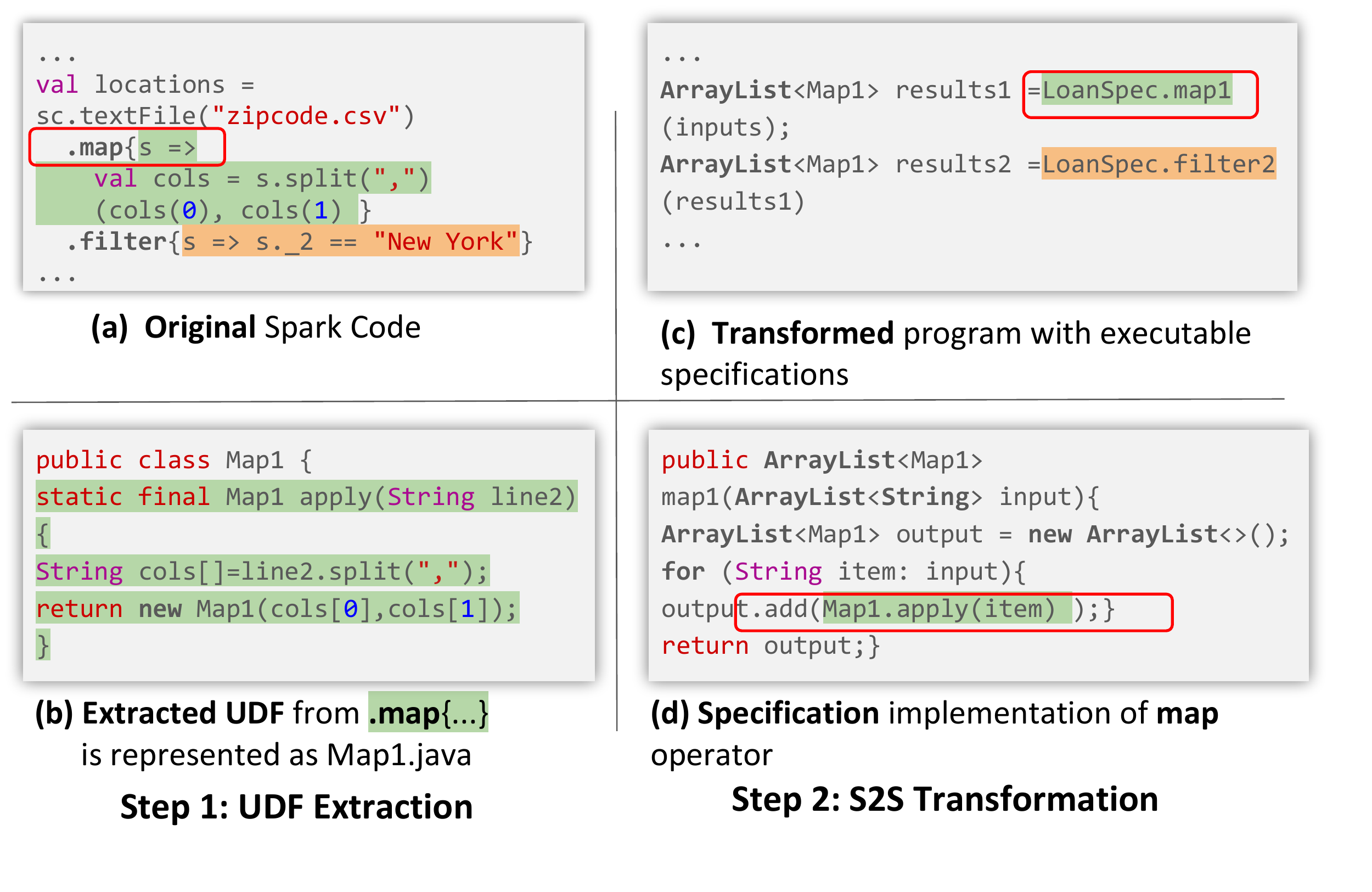}
\end{minipage}
\vspace{-2ex}
\caption{An Overview of \tool's Approach with Framework abstraction}\label{fig:bigfuzz}
\vspace{-2ex}
\end{figure*}

We devise a new coverage-guided, mutation-based fuzz testing approach for big data analytics called \tool. The key insight behind {\tool} is that {\em fuzz testing of DISC applications can be made tractable by abstracting framework code and analyzing application logic in tandem}. \tool first transforms a DISC application to a semantically equivalent, yet a framework-independent program that is more amenable to fuzzing to mitigate the long latency. It then uses a two-level instrumentation method to monitor control flow coverage of UDFs, while modeling the different outcomes of dataflow operations (i.e., dataflow equivalence classes). We call such a combination of behavior modeling as Joint Dataflow and UDF coverage (JDU coverage). During fuzzing, it uses schema-aware mutation operations guided by real-world error types to increase the chance of creating meaningful inputs that map to real-world errors.

The main contribution of \tool is that we made traditional fuzzing feasible for big data analytics by abstracting the dataflow behavior of the DISC framework with executable specifications. Additionally, \tool, to our knowledge, is the first fuzzing tool that uses mutations are specifically designed to reveal real-world DISC errors. In our experiments, \tool with framework abstraction can speed up the fuzzing time by 78 to 1477X compared to random fuzzing. Schema-aware mutation operations can improve application code coverage by 20 to 200\% with valid inputs as seeds, which leads to 33 to 100\% improvement in detecting application errors, when compared to naive random fuzzing. Even without valid input seeds, {\tool} improves application code coverage by 118 to 271\% and error detection by 58 to 157\%, demonstrating its robustness. 

The full technical paper on this approach appeared at ASE 2020~\cite{zhangbigfuzz} and this paper describes \tool's user interfaces and internal implementation with a focus on tool demonstration. \tool is a Java-based command line tool for testing Scala/Java Spark applications and can be easily generalized to other DISC frameworks such as Hadoop MapReduce~\cite{dean2004mapreduce}. We provide access to artifacts of  
{\tool} at 
\url{https://github.com/qianzhanghk/BigFuzz}.


\section{Technical Approach}\label{sec:approach}
We describe \tool's main technical contributions below. The detailed explanation is described in our full paper~\cite{zhangbigfuzz}. \tool takes in an Apache Spark program written in Scala or Java and an input schema, and it automatically generates concrete data for effective and efficient testing. Figure~\ref{fig:bigfuzz} illustrates the three novel parts in \tool.

\subsection{Framework Abstraction}
\tool maps each dataflow operator's implementation to a corresponding simplified yet semantically-equivalent implementation, which we call {\em executable specifications}. Such specifications help eliminate the dependency on the framework's code, transforming a DISC application into an equivalent, simplified Java program that can be invoked numerous times in a fuzzing loop. \tool automates this process in two steps: {\em UDF Extraction} and {\em Source-to-Source Transformation}.

In the first step, \tool decomposes the input Spark program into a direct acyclic graph (DAG) of dataflow operators and a list of corresponding UDFs. It decompiles the bytecode of the original Spark program into Java source code and traverses the Abstract Syntax Tree (AST) to extract each UDF into a separate Java class, as shown in Figure~\ref{fig:bigfuzz}b. It then uses the extracted DAG and UDFs to reconstruct the DISC application in the same interconnected dataflow order using executable specifications. For example in Figure~\ref{fig:bigfuzz}c, \codefont{map} operator is followed by \codefont{filter} operator, emulating their connection in Figure~\ref{fig:bigfuzz}a. The dataflow spec implementation, such as the spec of \codefont{map} operator in Figure~\ref{fig:bigfuzz}d, takes in an ArrayList object as input, applies the corresponding UDF on each element of the input list, and returns an output ArrayList.

\subsection{Coverage Guidance}
To differentiate UDFs from framework code, \tool designs a two level instrumentation and monitoring method for application specific coverage guidance. For dataflow operators, it monitors at the level of equivalence classes by extending the \codefont{TraceEvent} in JQF~\cite{JQF} to a specific \codefont{DataFlowEvent}. In addition to an identifier, \codefont{DataFlowEvent} has an additional Boolean or Integer variable to keep track of which subset of equivalence classes is exercised by the corresponding dataflow operator. For example, \codefont{"FilterEvent(arm=1)"} for \codefont{filter} operator represents the non-terminating equivalence class, where the filter predicate holds true and individual data records thus pass through the filter predicate. \codefont{"FilterEvent(arm=0)"} indicates the other terminating case, where the filter predicate holds false. For UDFs coverage, \tool uses a selective instrumentation scheme in ASM~\cite{asmtoolkit}, while ignoring all other dependent libraries. This combination of monitoring dataflow equivalence coverage together with control flow events in UDFs constitutes the JDU coverage, which essentially represents the behavior of application logic.

\subsection{Mutation}
Instead of bit-level mutations, \tool uses a user-defined schema to perform {\em record-level schema-aware mutations}\textemdash modifying data with respect to the structured data types as well as value ranges. In the schema, a user can indicate the number of columns, data type, and data distribution for each column of the input data. Unlike random bit-level mutations that produce unnatural inputs, each of the schema-aware mutations mimics a real-world error type in DISC applications that may lead to program crashes or failures at runtime. To this extent, we extensively investigate DISC application errors posted on popular Q/A forums and code repositories. 

\section{Implementation}\label{sec:implementation}
\tool is a Java-based command line tool that provides efficient fuzz testing of DISC applications. \tool is built on top of JQF~\cite{JQF}, a Java-based fuzz testing framework that instruments Java bytecode on-the-fly as classes are loaded by the JVM. \tool requires a test driver to indicate the test class and test method. A test driver is a JUnit-style test class with \codefont{@RunWith(JQF.class)} annotation on the test class and \codefont{@Fuzz} annotation on the test method. A user can use the following command line to invoke \tool:

\vspace{1em}
\noindent
    \parbox{\columnwidth}{\codefont{BigFuzz/bin/jqf-bigfuzz -c .:\$(BigFuzz/scripts/classpath.sh) testclass testmethod <maxTrials>}
}
\vspace{1em}

Given an input Spark program, \tool first reads its bytecode and translates it to Java source code with Java Decompiler (JAD)~\cite{jad}. It parses the AST of the de-compiled Java source code to search for a method invocation corresponding to each dataflow operator. The input arguments of such method invocations represent the UDFs, which are stored as separate Java classes. Next, based on the transformed program, \tool inserts code, for example, lines 6-10 in Figure~\ref{fig:dfinstru}, to each specification-based implementation of dataflow operators to monitor which equivalence class is activated, and it instruments the bytecode of the extracted UDF classes only to collect exercised branches in current execution. All \codefont{DataflowEvent}s and \codefont{TraceEvent}s are emitted to a coverage logger. Then during fuzzing, \tool will either randomly mutate the seed input or randomly generate valid inputs followed by mutating such inputs to increase cumulative coverage.

\begin{figure}
\begin{minipage}{1\linewidth}
\begin{lstlisting}[style=MyCSmallStyle]
public ArrayList<Tuple3> filter1(ArrayList<Tuple3> 
  input){
    ArrayList<Tuple3> ans = new ArrayList<>();
    for(Tuple3 item: input){
    if(Filter1.apply(item)) ans.add(item);}
    int iid = LoanSpec.class.hashCode();
    int arm = 0;
    if(ans.isEmpty()==false) arm=1;
    !\colorbox{white!70!orange}{TraceLogger.get().emit(new FilterEvent}!
    !\colorbox{white!70!orange}{(iid,arm));}!
    return ans;} 
\end{lstlisting}
\vspace{-1.5mm}
\end{minipage}
\vspace{-3mm}
\caption{Instrumented \codefont{filter} to emit dataflow equivalence class coverage} \label{fig:dfinstru}
\vspace{-2ex}
\end{figure}
\begin{figure}
\begin{minipage}{1\linewidth}
\begin{lstlisting}[style=MyCSmallStyle]
val data = text.map {
    s =>
    val cols = s.split(",")
    (cols(0), Integer.parseInt(cols(1)), Integer.parseInt(cols(2)))
    }.filter( s => s._1.equals("90024"))
val pair = data.map {
    s =>
    if (s._2 >= 40 & s._2 <= 65) {
        ("40-65", s._3)
    } else if (s._2 > 20 & s._2 < 40) {
        ("20-39", s._3)
    } else if (s._2 < 20){
        ("0-19", s._3)
    } else {
        (">65", s._3)
    }
}
val sum = pair.mapValues( x =>(x,1))
.reduceByKey((x, y) => (x._1 + y._1, x._2 + y._2))
.mapValues(x=>(x._2,x._1.toDouble/x._2.toDouble))
.foreach(println)

\end{lstlisting}
\end{minipage}
\caption{Alice's program that finds the average income per age group in her district.} \label{fig:sparkprogram}
\vspace{-2ex}
\end{figure}

\section{Demonstration}\label{sec:demo}
In this section, we present a step-by-step demonstration of \tool. Suppose Alice would like to investigate the average income per age range in her district. She uses the entire income survey database which contains the income information of states and counties for over several years. A sample row in this dataset is a string that contains the zipcode of employee, the age, and the annual income amount of this employee respectively (\eg  \fbox{\begin{minipage}[t][0.5em][t]{7em}\centering
\centerline{\texttt{\footnotesize{90095,33,58000}}} 
\end{minipage}}).

Alice writes a Spark application to perform this analysis, as shown in Figure~\ref{fig:sparkprogram}. She first uses a \codefont{map} operator to extract the zipcode, the age, and the income amount from each row using a UDF in line 4, and uses a \codefont{filter} operator to filter the data rows based on if its zipcode is "90024" in line 5. Next, Alice uses another \codefont{map} operator to cluster the data into different age groups. In lines 18-19, she aggregates all the income and number of persons in each age group with a \codefont{reduceByKey} operator. In the end, Alice calculates the average income with a \codefont{mapValues} operator in line 20.

To run \tool on her program, she writes an input schema shown in the following code snippet to describe her data. 
\fbox{\begin{minipage}[t][0.5em][t]{24em}\centering
\centerline{\texttt{\footnotesize{number string[900xx],integer[0-120],integer[0-$2^{32}$]}}} \end{minipage}} 
This indicates that each input entry comprises of three comma-separated columns: the first column must be a 5-bit number string with prefix "900", the second column must be an integer within the range [0-120], and the last column is an integer within [0-$2^{32}$]. \tool takes such input schema through \codefont{conf} file and passes it to \codefont{MutationGeneration} to automatically generate mutations that are tailored for this schema.

As a first step, \tool decomposes her program into six Java classes representing the UDFs, each of which can be identified with the operator name followed by its execution order (\eg, \codefont{map1.java}). Next, \tool reconstructs her program with these Java classes using the executable specifications and automatically generates a test driver for her program. Alice invokes \tool by typing the following command:

\vspace{1em}
\noindent
    \parbox{\columnwidth}{\codefont{BigFuzz/bin/jqf-bigfuzz -c .:\$(BigFuzz/scripts/classpath.sh) IncomAggregationDriver IncomeAggregation}
}
\vspace{1em}

Alice can monitor the fuzzing process by observing the console log and interrupts the execution by pressing "Ctrl-C". \tool does not require valid inputs as seeds. For each iteration, \tool produces a new input with several data rows by either randomly mutating the seed inputs or randomly generating valid inputs. Such new input will be saved to a separate \codefont{csv} file if it detects a unique crash or a new JDU branch.

Alice uses the generated test data 
\fbox{\begin{minipage}[t][0.5em][t]{7em}\centering
\centerline{\texttt{\footnotesize{90024,20,10900}}} 
\end{minipage}}
as input for unit testing. This input represents the income for a 20-year old person; however, the test output classifies it to \codefont{(>65)}. Alice investigates her code and identifies an error where a $\geq$ is misused by $>$ in line 10 of Figure~\ref{fig:sparkprogram}). \tool also generates inputs such as non-numeric strings or non-integer numbers to reveal critical runtime crashes in line 3 of Figure 2. Alice inserts relevant exception handling and data filter to eliminate such corner cases.


\section{Related Work}\label{sec:relatedwork}
\noindent{\bf Testing DISC Applications.}
Gulzar et al.~model the semantics of these operators in first-order logical specifications alongside the symbolic representation of UDFs~\cite{BigTest} and generate a test suite to reveal faults. Prior DISC testing approaches either do not model the UDF or model the specifications of dataflow operators partially~\cite{li2013sedge,Olston2009sigmod}. Li et al.~propose a combinatorial testing approach that automatically extracts input domain information from schema and bounds the scope of possible input combinations~\cite{bittag}. However, all these symbolic executions use a heuristic (loop iteration bound K) during path exploration, which may lead to false negatives, and they are also limited in applicability due to their symbolic execution scope. 

\noindent\textbf{Fuzz Testing.} 
Fuzz testing mutates the seed inputs through a \emph{fuzzer} to maximize a specific guidance metric, such as branch coverage, and find crashes in programs and frameworks. 
Fuzz testing has been shown to be highly effective in revealing a diverse set of bugs, including correctness bugs~\cite{babic2019fudge,afl}, security vulnerabilities~\cite{brennan2020jvm,gan2018collafl}, and performance bugs~\cite{wen2020memlock}. For example, AFL~\cite{afl} mutates a seed input to discover previously unseen coverage profiles. MemLock~\cite{wen2020memlock} employs both coverage and memory consumption metrics to find abnormal memory behavior.

Instead of flipping several bits/bytes in each mutation, ~\cite{lin2015browser} has investigated specific mutations for web browsers. \tool, along with our full technical paper~\cite{zhangbigfuzz}, designs mutations based on an empirical study of Apache Spark application errors reported in StackOverflow and Github. 
To speed up test execution while fuzzing, UnTracer~\cite{nagy2019full} dynamically strips out code-coverage instrumentation for lines of code that have already been covered. For DISC applications, the overhead is not due to instrumentation but indeed due to the extensive framework code. \tool is the first fuzzing tool that transforms the target application by simplifying framework logic. 

\section{Conclusion}\label{sec:conclusion}
To adapt fuzzing to DISC applications with long latency, we propose \tool that leverages (1) dataflow abstraction using source-to-source transformation, (2) tandem monitoring of equivalence-class based dataflow coverage with control flow coverage in user-defined functions, and (3) schema-aware mutations that reflect real world error types. \tool achieves up to 1477X speed-up compared to random fuzzing, improves application code coverage by up to 271\%, leading to up to 157\% improvement in detecting application errors.
\vspace{3pt}


\section{Acknowledgment} The participants of this research are in part supported by NSF grants CHS-1956322 CCF-1764077, CCF-1723773, ONR grant N00014-18-1-2037, and Intel CAPA grant.

\bibliographystyle{ieeetr}
\bibliography{reference}

\end{document}